\documentclass[twocolumn, 10pt, pre, aps, showpacs, reprint, amsmath, amssymb, superscriptaddress, nofootinbib]{revtex4-1}

\usepackage[pdftex]{graphicx}

\newcommand{\neff}{n}
\newcommand{\reffig}[1]{figure~\ref{#1}}
\newcommand{\Qmax}{Q_\mathrm{max}}
\newcommand{\taucl}{\tau_\mathrm{cl}}
\newcommand{\Qcl}{Q_\mathrm{cl}}
\newcommand{\chicrit}{\chi_\mathrm{cr}}
\renewcommand{\Re}{\mathrm{Re}}
\renewcommand{\Im}{\mathrm{Im}}

\begin{document}

\title{Spatial structure of lasing modes in wave-chaotic semiconductor microcavities}

\author{Stefan Bittner}
\affiliation{Department of Applied Physics, Yale University, New Haven, Connecticut 06520, USA}
\affiliation{Chair in Photonics, LMOPS EA-4423 Laboratory, CentraleSup\'elec and Universit\'e Paris-Saclay, 2 rue Edouard Belin, Metz 57070, France}
\author{Kyungduk Kim}
\affiliation{Department of Applied Physics, Yale University, New Haven, Connecticut 06520, USA}
\author{Yongquan Zeng}
\author{Qi Jie Wang}
\affiliation{Center for OptoElectronics and Biophotonics, School of Electrical and Electronic Engineering and the Photonics Institute, Nanyang Technological University, 639798 Singapore}
\author{Hui Cao}
\email{Corresponding author: hui.cao@yale.edu}
\affiliation{Department of Applied Physics, Yale University, New Haven, Connecticut 06520, USA}

\begin{abstract}
We present experimental and numerical studies of broad-area semiconductor lasers with chaotic ray dynamics. The emission intensity distributions at the cavity boundaries are measured  and compared to ray tracing simulations and numerical calculations of the passive cavity modes. We study two different cavity geometries, a D-cavity and a stadium, both of which feature fully chaotic ray dynamics. While the far-field distributions exhibit fairly homogeneous emission in all directions, the emission intensity distributions at the cavity boundary are highly inhomogeneous, reflecting the non-uniform intensity distributions inside the cavities. The excellent agreement between experiments and simulations demonstrates that the intensity distributions of wave-chaotic semiconductor lasers are primarily determined by the cavity geometry. This is in contrast to conventional Fabry-Perot broad-area lasers for which the intensity distributions are to a large degree determined by the nonlinear interaction of the lasing modes with the semiconductor gain medium. 
\end{abstract}

\maketitle

\section{Introduction}
Broad-area semiconductor lasers are commonly employed for high-power applications such as machining, material processing or medical surgery. The typical geometry is a Fabry-Perot cavity with broad cross section of the order of $100~\mu$m, which is necessary to achieve high powers but leads to lasing in several spatial (transverse) modes. The emission intensity distributions are not simply determined by the passive cavity resonances since the nonlinear interactions of the optical field with the gain medium lead to lensing and self-focusing that create spots of high intensity, so-called filaments \cite{Mehuys1987, Abraham1990, Lang1991}. Since the filaments are intrinsically unstable, the lasing emission patterns exhibit spatio-temporal fluctuations on a sub-nanosecond timescale \cite{Fischer1996, Hess1996, Marciante1997, OhtsuboBook2013}. Intensive efforts \cite{Takimoto2009, Simmendinger1999, Mandre2005, Gehrig1999, Adachihara1993} have been made to stabilize the lasing dynamics because a temporally stable beam profile is required for many applications. 

Recently it has been shown that broad-area semiconductor lasers with D-shaped cavities can suppress the spatio-temporal instabilities from which the conventional Fabry-Perot type broad-area lasers suffer \cite{Bittner2018a}. In contrast to the regular ray dynamics in a Fabry-Perot cavity, the D-shaped cavity features fully chaotic ray dynamics. Instead of propagating mainly along one axis as in a Fabry-Perot cavity, the rays in the D-shaped cavity travel into all possible directions. Following the principle of ray-wave correspondence, the resonant modes consist of plane wave components with all possible propagation directions, and the resulting complex interference prevents self-focusing and filamentation. However, the lasing emission intensity distributions on the cavity boundary are very inhomogeneous with regions of high as well as very low intensity \cite{Bittner2018a}. This experimental observation raises the question to what extent the structure of the lasing modes is influenced by the asymmetric cavity geometry and the nonlinear light-matter interaction, respectively. 

Asymmetric dielectric microcavities have been intensively studied for laser applications \cite{Tureci2005, Harayama2010, Xiao2010b, Cao2015}. Most asymmetric resonators feature at least partially chaotic ray dynamics and are hence called wave-chaotic cavities. Dielectric resonators are leaky systems because rays can escape refractively, and their properties consequently differ significantly from those of closed cavities. Semiclassical methods \cite{Brack2003} and ray tracing simulations \cite{Altmann2013, Cao2015} have proven very effective to understand and predict their spectral properties and emission directions. Most studies concentrate on the far-field intensity distributions and the phase space representations of the modes (so-called Husimi distributions \cite{Husimi1940, Hentschel2003}) instead of the intensity distributions inside the cavities or at the cavity boundaries. 

Here we focus on the lasing intensity distributions inside the fully chaotic dielectric microcavities and at the cavity boundaries. The degree of spatial localization of lasing modes determines the strength of modal competition for gain and thus the number of lasing modes, with important consequences for the spatial coherence of the emission \cite{Cao2019, Cerjan2019}. Although the intra-cavity intensity distributions cannot be easily measured experimentally, the emission profiles at the cavity boundaries allow to draw conclusions about the spatial structure of the lasing modes inside asymmetric cavities \cite{Lafargue2014, Bittner2016, Bittner2018, Alekseev2018}. Furthermore, knowing the locations of intense emission at the cavity boundary enables efficient coupling into a local waveguide. 

Our aim is to understand the roles that the cavity geometry and the nonlinear modal interactions play in determining the lasing intensity distributions of wave-chaotic cavities. We fabricate and investigate GaAs quantum well lasers with two different cavity shapes, D-cavity and stadium, both featuring fully chaotic ray dynamics. Although the D-cavities and stadia emit fairly homogeneously in all directions, the intensity distributions inside the cavities and at the cavity boundaries are very inhomogeneous above lasing threshold. Furthermore the coarse structure (i.e., envelope) of the emission intensity distributions at the cavity boundaries is independent of the pump current above lasing threshold and scales with the cavity size, which indicates that the coarse structure is dictated by the cavity shape rather than by the nonlinear light-matter interactions. This is additionally confirmed by numerical calculations of the passive cavity modes and ray tracing simulations which show that the structure of the intensity distributions results from refractive escape of light from the cavity and is completely determined by the passive cavity modes with high quality ($Q$) factor. Moreover, the excellent agreement with ray tracing simulations demonstrates that the principle of ray-wave correspondence \cite{Tureci2005, Cao2015} holds for fully chaotic cavities even in the presence of nonlinear interactions between lasing modes and gain medium. The ray tracing calculations accurately predict not only the intensity distributions inside and outside of the cavities but also the quality factors of the most long-lived modes. Such predictions are particularly valuable for cavities that are much larger than the wavelength and are thus not accessible for wave simulations. 

The article is organized as follows. In Section~\ref{sec:exp}, we present the experiments and the measurement results. Section~\ref{sec:sim} describes the wave and ray simulations, and in Section~\ref{sec:comp} we compare the results of wave calculations, ray tracing and experimental measurements. We conclude with a summary and outlook in Section~\ref{sec:conc}. 

\section{Experiments} \label{sec:exp}

\begin{figure}[tb]
\begin{center}
\includegraphics[width = 8.4 cm]{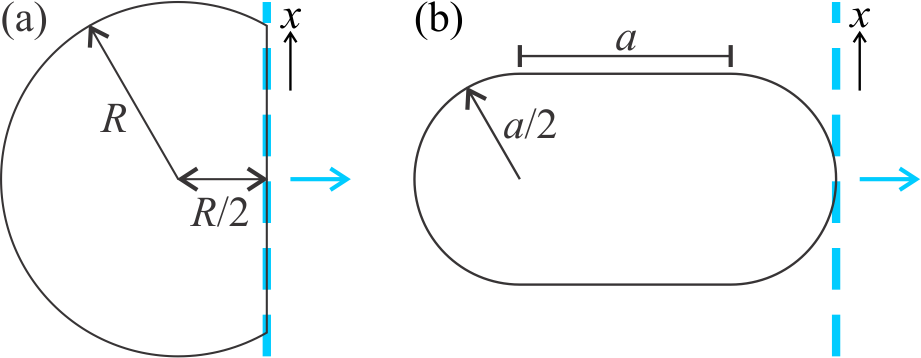}
\end{center}
\caption{Geometries of (a) D-cavity with radius $R$ and cut $R/2$ away from the center and (b) stadium consisting of a square with side length $a$ between two semicircles with radius $a/2$. The dashed blue lines indicate the image planes for the sidewall emission intensity distributions ($x$-axis) and the blue arrows indicate the directions in which the emission is observed.}
\label{fig:cavGeom}
\end{figure}

The edge-emitting semiconductor microlasers are fabricated from a commercial GaAs/AlGaAs quantum well epiwafer (Q-Photonics QEWLD-808) with photolithography and inductively coupled plasma dry etching (see \cite{Bittner2018a} for details). The etching depth is about $4~\mu$m to ensure a strong refractive index contrast at the cavity boundary for good optical confinement. The effective refractive index of the cavity is $\neff = 3.37$. 

We investigate two types of wave-chaotic microcavities. The first one is a D-cavity, which is a circle with a segment cut off [see \reffig{fig:cavGeom}(a)]. A D-cavity larger than a semicircle has completely chaotic ray dynamics \cite{Bunimovich1979, Ree1999}. Here we consider the D-cavity with the cut $R/2$ away from the center of the circle with radius $R$ because the average Lyapunov exponent of the ray trajectories is approximately the largest for this geometry. The second cavity has the shape of a stadium, which comprises a rectangle between two semicircles [see \reffig{fig:cavGeom}(b)]. The ray dynamics in a stadium is completely chaotic as well \cite{Bunimovich1979}, and we consider the stadium with a square between the semicircles so that the average Lyapunov exponents of the ray trajectories is approximately maximized. Experimentally, D-cavities with radii $R = 100$ and $200~\mu$m as well as stadia with $a = 119~\mu$m are investigated, where the stadia with $a = 119~\mu$m have the same area as the D-cavities with $R = 100~\mu$m (cf.\ \ref{sec:cavArea}). 

The microlasers are pumped electrically with $2~\mu$s-long pulses at a repetition rate of $10-50$~Hz to reduce heating. All experiments are performed at ambient temperature. A $20\times$ microscope objective ($\mathrm{NA} = 0.40$) is used to collect the emission from one of the cavity sidewalls. The emission is coupled into a multimode fiber bundle connected to a spectrometer for measuring the lasing spectrum. For spatial measurements, the objective is used in conjunction with a second lens with $f = 150$~mm in a $2f$-$2f$ configuration to image the emission intensity distributions on the sidewalls on a CCD camera (Allied Vision Mako G125-B, see \cite{Bittner2018a} for more details of the setup). The image planes used for D-cavities and stadia are indicated by the blue dashed lines in \reffig{fig:cavGeom}. A long working-distance objective ($\mathrm{NA} = 0.42$) is used to make top view images of the lasers with a second CCD camera (Allied Vision Mako G234-B). 

\begin{figure*}[tb]
\begin{center}
\includegraphics[width = 16 cm]{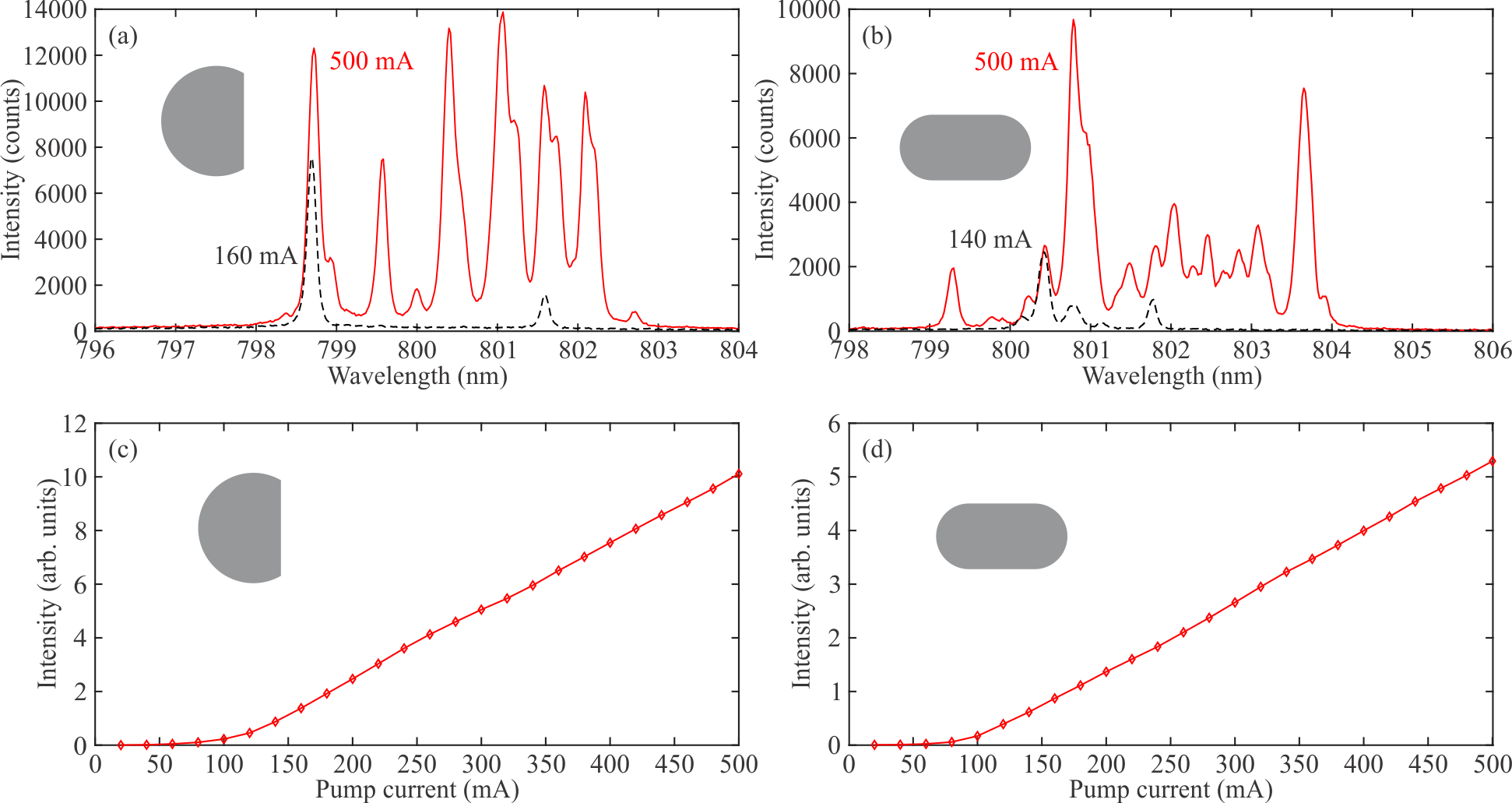}
\end{center}
\caption{Lasing spectra integrated over a single $2~\mu$s-long current pulse and LI-curves (light intensity $L$ vs pump current $I$). (a)~Spectra of a D-cavity with $R = 100~\mu$m at pump currents $160$~mA (black dashed line) and $500$~mA (red solid line). (b)~Spectra of a stadium with $a = 119~\mu$m at pump currents of $140$~mA (black dashed line) and $500$~mA (red solid line). (c)~LI-curve of the D-cavity with threshold current $I_{th} = 130$~mA (threshold current density $j_{th} = 514~\mathrm{A~cm}^{-2}$). (d)~LI-curve of the stadium cavity with $I_{th} = 100$~mA ($j_{th} = 396~\mathrm{A~cm}^{-2}$).}
\label{fig:spectraLI}
\end{figure*}

Figures~\ref{fig:spectraLI}(a) and \ref{fig:spectraLI}(b) show typical lasing spectra of a D-cavity and stadium, respectively. The lasers operate in a multimode regime even close to threshold. It is interesting to note that single-mode lasing for stadium-shaped semiconductor microlasers was observed by other groups \cite{Sunada2016a}; see \cite{Cerjan2019} for a detailed discussion. The polarization of the laser emission is purely transverse electric (TE, electric field parallel to the plane of the cavity). The light-current (LI) curves in figures~\ref{fig:spectraLI}(c) and \ref{fig:spectraLI}(d) show a clear threshold, which is at about $I_{th} = 130$~mA ($100$~mA) for the D-cavity (stadium). These values of the threshold currents are typical and are confirmed for multiple cavities. The slopes of the LI-curves depend on the collection efficiency of the objective and can hence not be compared quantitatively. 

\begin{figure}[tb]
\begin{center}
\includegraphics[width = 8.4 cm]{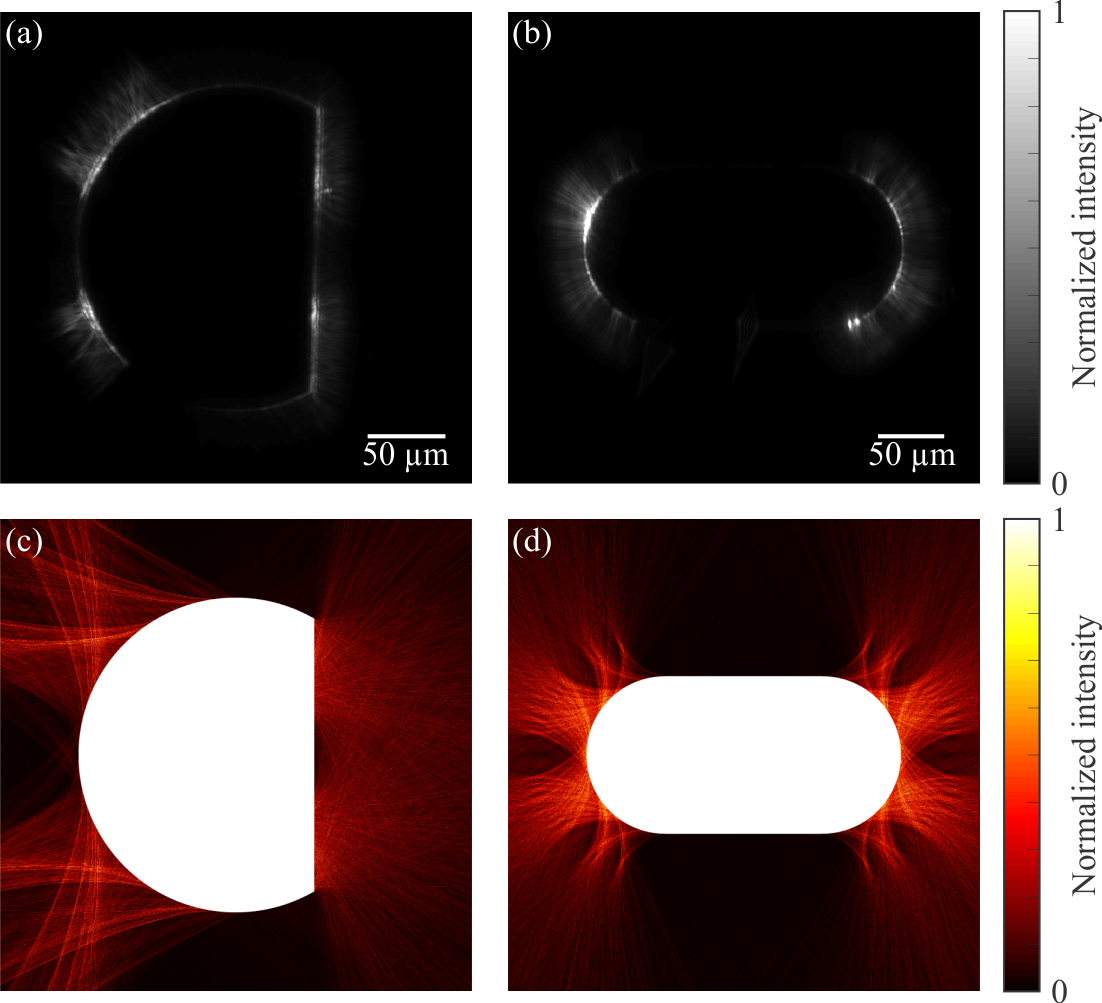}
\end{center}
\caption{Top-view optical microscope images of (a) a D-cavity with $R = 100~\mu$m and (b) a stadium with $a = 119~\mu$m. Part of the lasing emission at the cavity sidewall is scattered out of plane and towards the camera on top by the roughness of the substrate. The pump current is $500$~mA and the images are integrated over a single $2~\mu$s-long pump pulse. (c) Ray tracing simulations of the emitted light intensity distributions just outside of a D-cavity and (d) a stadium.}
\label{fig:topViewNF}
\end{figure}

The top view microscope images of the cavities in figures~\ref{fig:topViewNF}(a) and \ref{fig:topViewNF}(b) show the lasing emission that is diffracted towards the substrate and then scattered in the vertical direction by small scattering centers on the substrate in the vicinity of the cavities. The images indicate that the emission intensity distributions of the cavity are very inhomogeneous and exhibit the same mirror symmetries as the cavities. For example, the D-cavity shows almost no emission from the middle of its straight sidewall, whereas its top and bottom third feature strong emission. For the stadium, almost the complete emission originates from the semicircular parts of the boundary, whereas the emission from the straight sidewalls is negligible. 

\begin{figure*}[tb]
\begin{center}
\includegraphics[width = 12 cm]{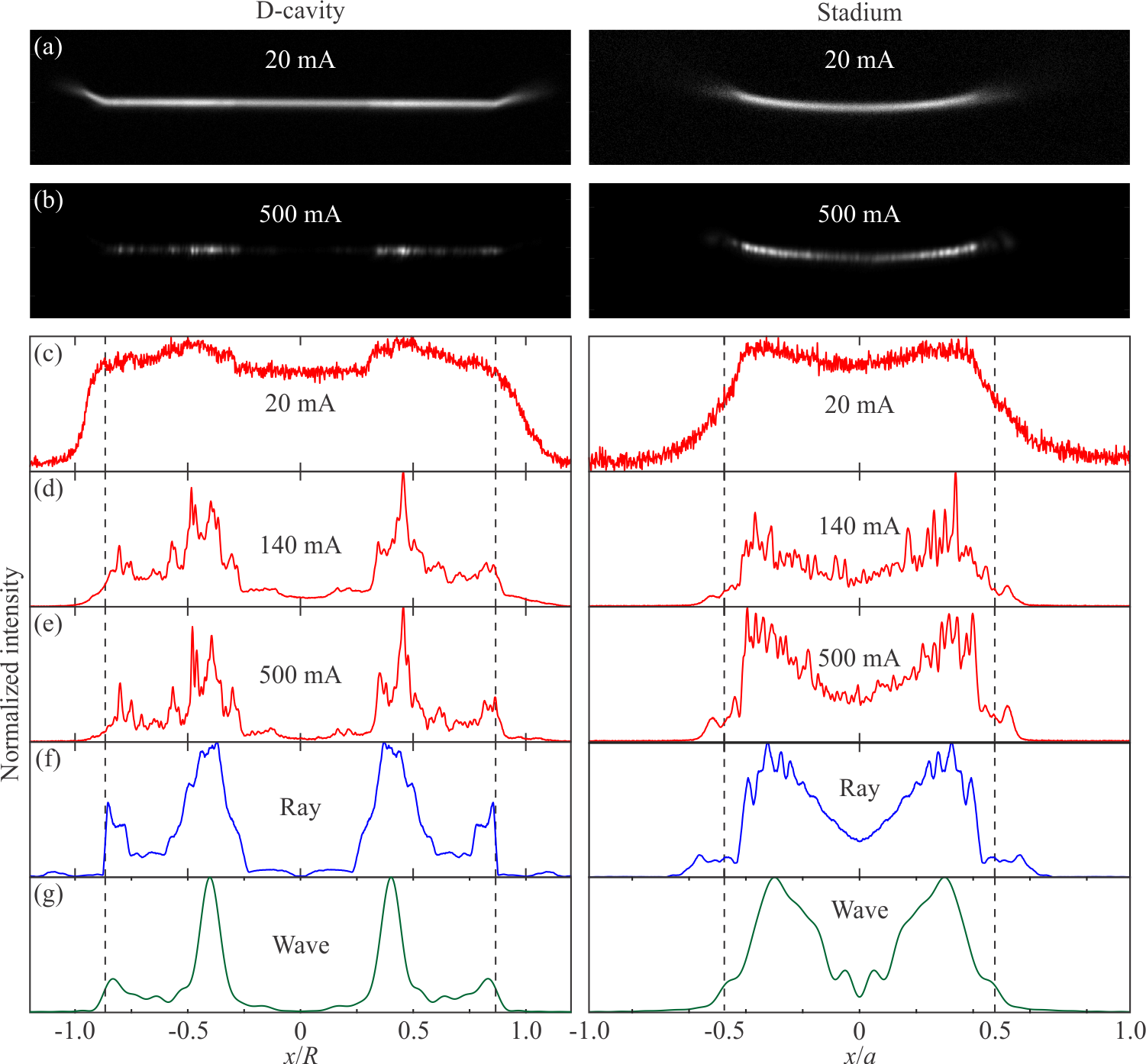}
\end{center}
\caption{Emission intensity distributions on the sidewalls of a D-cavity with $R = 100~\mu$m (left column) and a stadium with $a = 119~\mu$m (right column) integrated over a single $2~\mu$s-long pump pulse. (a)~CCD images well below threshold ($20$~mA) and (b)~well above ($500$~mA). (c)~Measured emission intensity distributions for $20$~mA, (d)~$140$~mA, and (e)~$500$~mA. (f)~Ray tracing simulations of the emission intensity distributions for collecting $\mathrm{NA} = 0.4$. (g)~Wave simulations of the emission intensity distributions with collecting $\mathrm{NA} = 0.4$. The intensity distributions of $11$ ($24$) high-$Q$ modes of a D-cavity with $R = 20~\mu$m (stadium with $a = 23.8~\mu$m) are added. The vertical dashed lines mark the locations of the intersections of the straight segments and the circular boundaries at $x = \pm \sqrt{3} R / 2$ ($x = \pm a / 2$) for the D-cavity (stadium).}
\label{fig:sweid}
\end{figure*}

The straight sidewall of a D-cavity and the plane touching the semicircle of a stadium (blue dashed lines in \reffig{fig:cavGeom}) are imaged onto a CCD camera to enable a more quantitative measurement of the emission intensity distributions. Figure~\ref{fig:sweid}(a) shows the CCD images of a D-cavity and a stadium pumped well below threshold. Both cavities feature a fairly homogeneous emission intensity distribution. When pumped above threshold, however, the emission profiles of both cavities become very inhomogeneous as shown in \reffig{fig:sweid}(b). Figures~\ref{fig:sweid}(c)--(e) show the emission distributions obtained from the CCD images by integrating in the direction perpendicular to the cavity plane. Figure~\ref{fig:sweid}(c) shows that the emission profiles below threshold are in fact not completely homogeneous, but have very little variation along $x$. However, already just above threshold, the emission distributions are very inhomogeneous as shown in \reffig{fig:sweid}(d). The emission intensity distributions above threshold feature two different length scales: sharp peaks with widths of the order of $1~\mu$m, and a coarse structure (i.e., envelope) that varies on a length scale of several $10~\mu$m. Most notably, the emission distributions exhibit a region of low intensity in the middle of the sidewall for both D-cavity and stadium, a feature which already starts to develop below the lasing threshold [see \reffig{fig:sweid}(c)]. 

When increasing the pump, the fine features of the emission intensity distributions change, whereas the coarse structure stays the same as shown in \reffig{fig:sweid}(e). The intensity distribution of a lasing mode in a wave-chaotic cavity evidently features variations on the scale of the wavelength which results in the sharp peaks. In the experiments, however, their width is limited by the finite numeric aperture of the objective, $\mathrm{NA} = 0.40$, which yields the length scale of $1~\mu$m. The differences in the fine structure that appear with increasing pump are due to the changes of the lasing modes and their relative intensities (cf.\ \reffig{fig:spectraLI}). The fact that the coarse structure does not change as a function of the pump current indicates that it is determined by a mechanism that is independent of specific lasing modes. 

\begin{figure}[tb]
\begin{center}
\includegraphics[width = 6.5 cm]{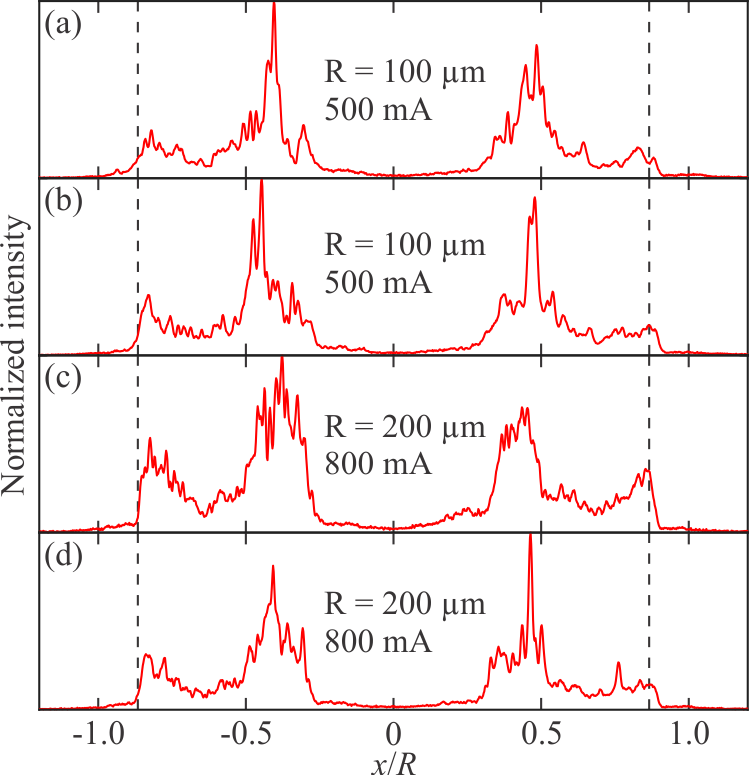}
\end{center}
\caption{Emission intensity distributions on the sidewalls of (a, b) two different D-cavities with $R = 100~\mu$m pumped with $500$~mA and (c, d) two different D-cavities with $R = 200~\mu$m pumped with $800$~mA. The vertical dashed lines indicate the two ends of the straight cut at $x = \pm \sqrt{3} R / 2$.}
\label{fig:sweidsDiffCavs}
\end{figure}

The sidewall emission intensity distributions of various D-cavities with different sizes ($R = 100$ and $200~\mu$m) are shown in \reffig{fig:sweidsDiffCavs}. All measured emission profiles exhibit the same coarse structure with a region of very low intensity in the middle flanked by regions of high intensity. Moreover, this coarse structure scales linearly with the cavity size so that the patterns match when plotted as a function of $x/R$ as in \reffig{fig:sweidsDiffCavs}. Analogous results are obtained for the stadia (not shown). While the measurements presented here are integrated over a single current pulse, time-resolved measurements (see \cite{Bittner2018a}) demonstrate that the same coarse structure is observed at any given time during a pulse with fluctuations of the fine structure only. 

\begin{figure*}[tb]
\begin{center}
\includegraphics[width = 15 cm]{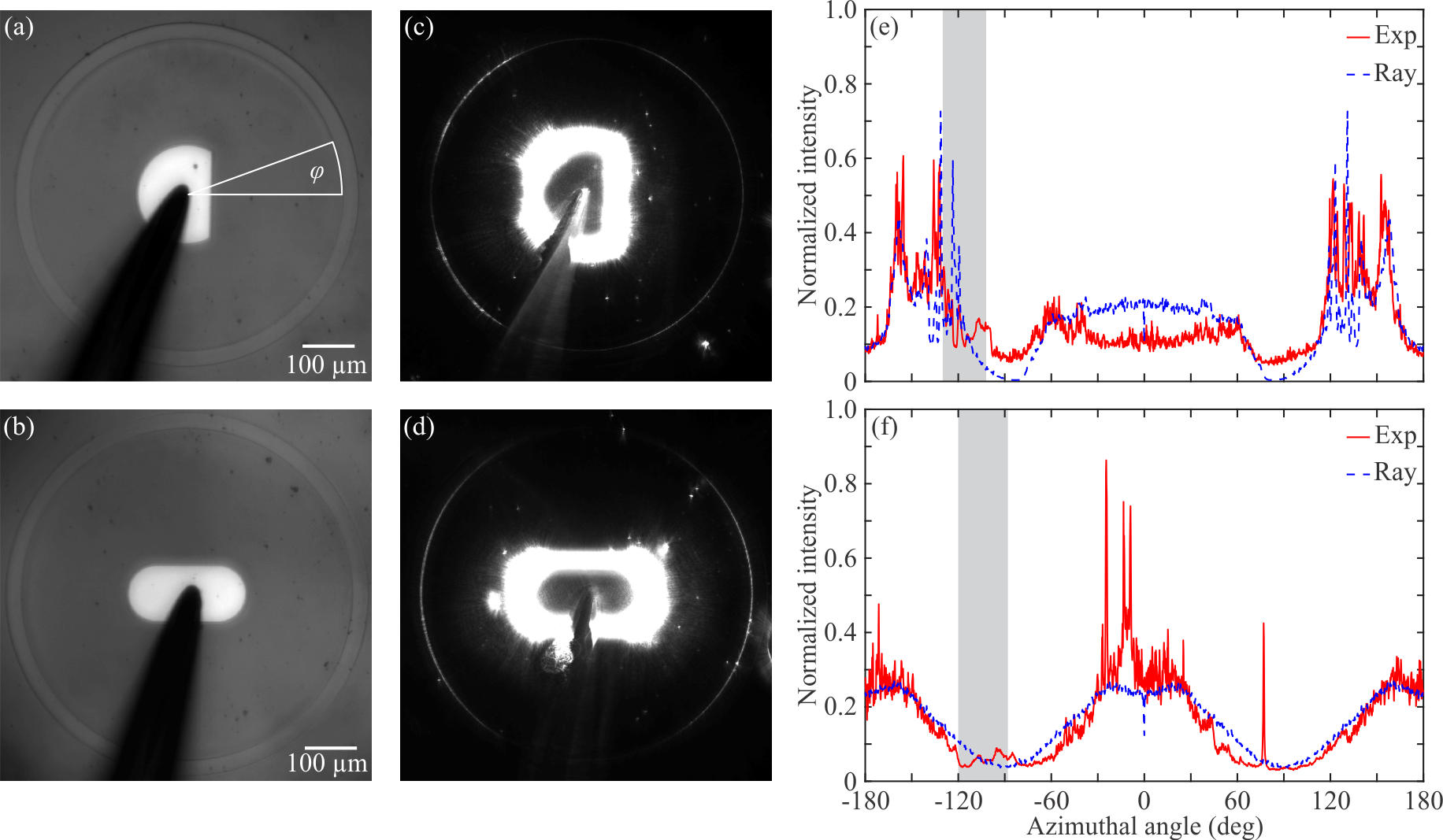}
\end{center}
\caption{Top-view optical images of lasing emission. (a)~A D-cavity with $R = 100~\mu$m is surrounded by a circular ring with radius $300~\mu$m and (b)~a stadium with $a = 119~\mu$m is surrounded by a ring with radius $319~\mu$m that scatters the light emitted from the cavity sidewalls towards the camera on top. (c)~Top view image of lasing emission from the D-cavity and (d)~the stadium at a pump current of $500$~mA. The tungsten needle that injects current into the microlaser casts a shadow and partially scatters the lasing emission. Due to the long exposure times necessary to record the light scattered by the ring, the camera is saturated by the emission scattered near the cavity boundaries. (e)~Measured emission intensity (red) at the inner boundary of the ring for the D-cavity as a function of the azimuthal angle $\varphi$ defined in (a) and (f)~for the stadium pumped with $500$~mA. The corresponding results from ray simulations are shown by blue dashed lines. The area below the curves is normalized to unity. The gray areas mark the regions covered by the needles where the signal is distorted. Additional distortions stem from scattering by small particles on the substrate and defects on the rings.}
\label{fig:MidFieldDistr}
\end{figure*}

In order to complement the information of the intensity distributions measured at the cavity sidewalls which show the origins of intense emission, the directions of emission are measured by scattering the light escaping from the cavities with a ring surrounding them at a distance. The ring has a radius of $300~\mu$m in the case of the D-cavity with $R = 100~\mu$m shown in \reffig{fig:MidFieldDistr}(a) and a radius of $319~\mu$m in the case of the stadium with $a = 119~\mu$m shown in \reffig{fig:MidFieldDistr}(b). Because the radius of the scattering rings is of the same order of magnitude as the cavity sizes for technical reasons, the scattering rings are \textit{not} in the far field. Figures~\ref{fig:MidFieldDistr}(c) and \ref{fig:MidFieldDistr}(d) shows the top view emission images of the D-cavity and the stadium pumped with $500$~mA, respectively. The image for the D-cavity (stadium) was integrated over $250$ ($300$) pump pulses, and hence the camera is saturated by the emission scattered directly near the cavity sidewalls (cf.\ \reffig{fig:topViewNF}). 

The emission intensity scattered at the outer ring is plotted as a function of the azimuthal angle $\varphi$ in figures~\ref{fig:MidFieldDistr}(e) and \ref{fig:MidFieldDistr}(f), respectively. The intensity distributions in \reffig{fig:MidFieldDistr} exhibit the symmetries expected from the cavity geometries, however, there are some perturbations due to the needles (indicated by the gray areas). Furthermore, the presence of scatterers and other defects near the rings leads to artificial peaks in the distributions, e.g., at $\varphi = 75^\circ$ and in the region $\varphi = -30^\circ$ to $0^\circ$ in \reffig{fig:MidFieldDistr}(f). For the D-cavity, the majority of the emission intensity is in the range of $\varphi = \pm (120^\circ$--$150^\circ)$ and in the range of $\varphi = -60^\circ$ to $+60^\circ$. For the stadium, the emission is concentrated in two broad regions around $0^\circ$ and $180^\circ$. Therefore, even though the emission from the cavities is not completely homogeneous in all directions as in the case of a circle cavity, it lacks strong directionality. It should be noted, however, that there are other cavity geometries with predominantly chaotic ray dynamics such as the lima\c{c}on that exhibit highly directional emission \cite{Wiersig2008}. Like the emission intensity distributions at the cavity boundaries in \reffig{fig:sweid}, the intensity distributions at the scattering rings show little dependence on the pump current above the lasing threshold.  

\section{Wave and ray simulations} \label{sec:sim}

The experimental observations in the previous section strongly indicate that the emission intensity distributions are determined not by nonlinear interactions but by the cavity geometry. First, the coarse structure of the emission intensity distributions does not depend on the pump current above the lasing threshold. Second, the same coarse structure is found for different cavities of the same size but different realizations of surface roughness. Third, the coarse structure scales linearly with the cavity size for the same resonator shape. If, in contrast, the structure of the intensity distributions resulted from the nonlinear interaction with the active medium, it would change with the pump current which increases the strength of the interaction. Moreover, the length scales of the structure would be mainly determined by the details of the light-matter interaction \cite{Marciante1997} rather than by the cavity size. 

In order to understand the structure of the lasing modes and how it is influenced by the cavity geometry, we compare the intensity distributions of lasing emission with calculations of the passive cavity modes. Because calculations of the passive modes are only feasible for cavity sizes significantly smaller than the experimental ones, we perform ray tracing simulations in order to obtain the relation between the cavity geometry and the structure of the intensity distributions in the semiclassical limit. This relation will be directly applicable to the experimental cavities as ray tracing simulations are scale free. 

\subsection{Simulations of passive cavity modes}
We calculate the passive cavity modes (also called resonances) of the D-cavities and stadia by solving the two-dimensional scalar Helmholtz equation
\begin{equation} [ \Delta + n^2(x, y) k^2 ] \Psi(x, y) = 0 \end{equation}
with outgoing wave boundary conditions \cite{Tureci2005}, where $\neff$ is the effective refractive index and $k = 2 \pi / \lambda$ is the wave number with $\lambda$ the free-space wavelength. Here the wave function $\Psi$ corresponds to the $z$-component of the magnetic field, $H_z$, as experimentally the lasing emission is TE polarized. The modes are calculated numerically using the Comsol eigenmode solver. The refractive index of the cavity is set to $\neff = 3.37$ and the free-space wavelength is around $800$~nm as in the experiments. 

Calculations are performed for a D-cavity with $R = 20~\mu$m, which corresponds to $kR \simeq 157$, and a stadium with $a = 23.8~\mu$m, which corresponds to $ka \simeq 187$. Both cavities have the same area, but have $5$--$10$ times smaller linear dimensions than those used in the experiments due to the restrictions of available computing power. However, they are quite far in the semiclassical limit $kR \gg 1$, and it was checked that simulations with cavities with half the linear dimensions yield the same qualitative results. Moreover, the wave calculation results agree well with the ray tracing simulations which confirms that they can be compared to the experimental data. 

\begin{figure*}[tb]
\begin{center}
\includegraphics[width = 16 cm]{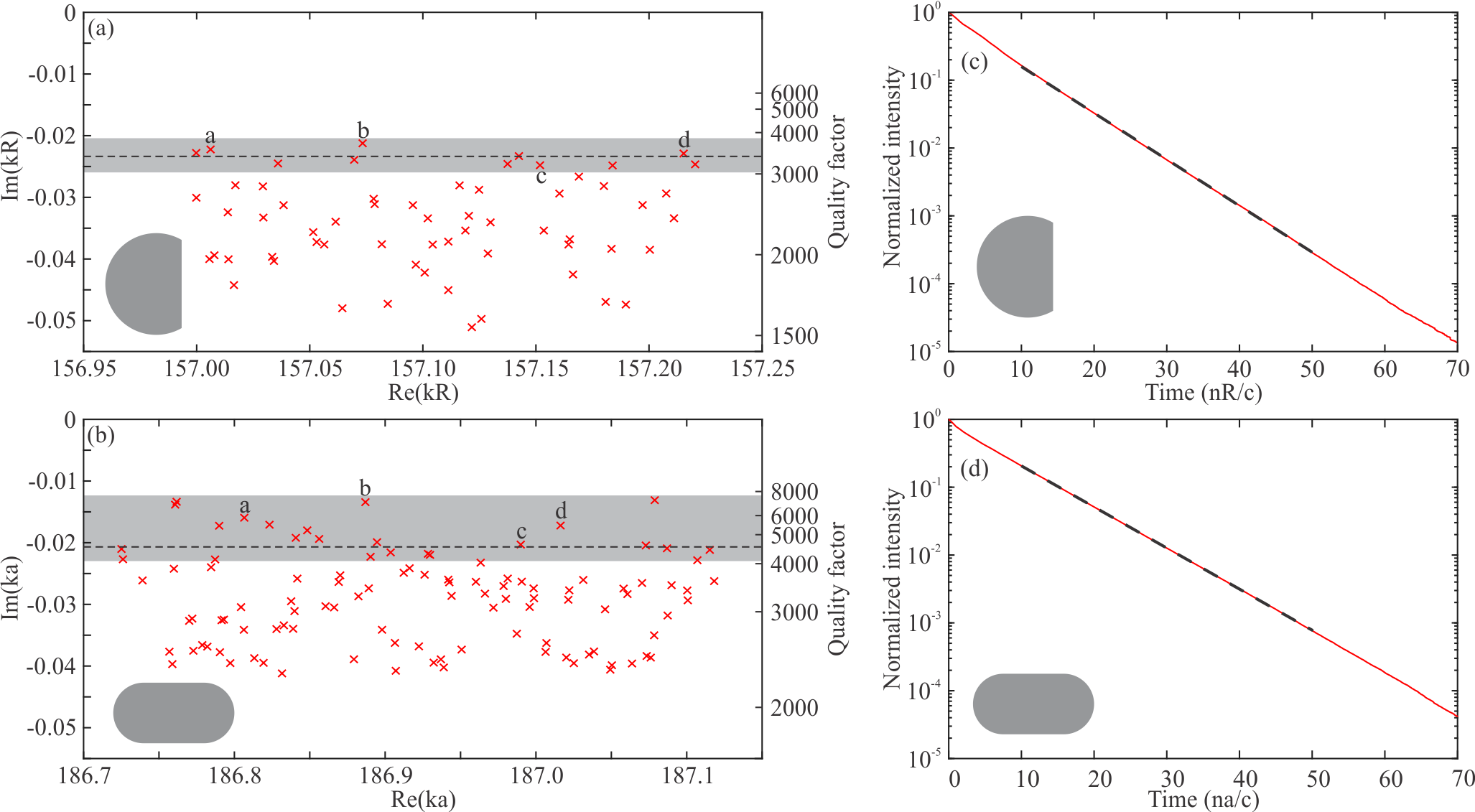}
\end{center}
\caption{(a,~b) Calculated resonant modes of the passive cavities with normalized complex frequencies given by $\Re(kR)$ and $\Im(kR)$ [$\Re(ka)$ and $\Im(ka)$] for the D-cavity [stadium]. (c,~d) Ray tracing simulations of the intensity decay of classical trajectories. The wave and ray simulations are for TE polarization and refractive index $\neff = 3.37$. The results for a D-cavity with $R = 20~\mu$m  are shown in panels (a,~c) and those for a stadium with $a = 23.8~\mu$m in panels (b,~d). The gray areas in the spectra indicate the modes with (a) $Q \geq 0.9~\Qcl = 3026$ and (b) $Q \geq 0.9~\Qcl = 4067$ used for the sum of intensity distributions discussed in the following. The intensity distributions of the modes labeled with a-d in the spectrum of the D-cavity (stadium) are shown in \reffig{fig:IID-Dcav} (\reffig{fig:IID-Stad}). The dashed black lines in the right panels are exponential fits with (c) $\taucl = 6.3507~\neff R / c$ and (d) $\taucl = 7.1736~\neff a /c$. The corresponding imaginary parts of the normalized frequencies, $\Im(kR)$ [$\Im(ka)$], are indicated as black dashed lines in the spectra (a, b).}
\label{fig:spectraDecay}
\end{figure*}

Figure~\ref{fig:spectraDecay} shows the calculated spectra, which consist of $61$ resonances for the D-cavity and $106$ for the stadium. Only the modes with the highest quality factors in the given wavelength range are calculated, whereas additional modes with lower $Q$-factors exist but are of no relevance here since they have no realistic chance of lasing. The most long-lived mode of the D-cavity has $\Qmax = 3699$ [$\Im(kR) = -0.0212$], whereas the modes of the stadium can have significantly higher $Q$-factors up to $\Qmax = 7137$ [$\Im(ka) = -0.0131$]. This raises the question what determines and limits the highest $Q$-factor for a given cavity and how $\Qmax$ depends on the cavity geometry, size and refractive index, which is hard to determine with wave simulations alone. 

\subsection{Ray dynamics simulations}
In order to answer these questions, we performed ray tracing simulations of dielectric billiards with D-cavity and stadium shape. The ray simulations allow to explore the semiclassical regime beyond the cavity sizes that can be treated with wave simulations, and moreover the comparison of wave and ray simulations enables us to distinguish between effects due to the classical ray dynamics and wave effects such as tunneling and scarring \cite{Heller1984}. As shown in the following, the ray simulations can explain both the intensity distributions and the different $Q$-factors (or lasing thresholds) of the D-cavities and stadia. 

For closed cavities, the semiclassical eigenfunction hypothesis states that the wave functions are supported by the regions of phase space explored by typical classical trajectories in the semiclassical limit, that is, when the cavity size is much larger than the wavelength \cite{Berry1977}. This implies that the average intensity distributions of resonators with fully chaotic (and thus ergodic) ray dynamics become uniform \cite{Shnirelman1974, ColindeVerdiere1985, Zelditch1996}. However, the dielectric resonators considered here are open systems since rays can escape by refraction at the dielectric interfaces that form the cavity sidewalls. Hence, in the classical limit of ray optics, they are leaking Hamiltonian systems \cite{Altmann2013}, which results in non-uniform intensity distributions. At the moment, the generalization of the semiclassical eigenfunction hypothesis to dielectric resonators remains an unsolved problem \cite{Claus2018, Claus2019}. 

Nonetheless, some properties of the wave functions are known for dielectric resonators with integrable or chaotic ray dynamics. For the integrable case, the resonant modes localize on classical tori \cite{Bittner2013b, Bittner2014a, Sukharevsky2019}. For chaotic ray dynamics, it has been shown that the modes with the highest $Q$-factors are based on a particular set of trajectories, the unstable manifold of the chaotic saddle \cite{Altmann2013, Casati1999b, Schwefel2004, Lee2004, Wiersig2008}. The chaotic saddle consists of the trajectories that stay confined forever in the cavity both for forward and backward propagation in time, and the unstable manifold is the set converging to the chaotic saddle for backward propagation\footnote{The unstable manifold is hence also called backward-trapped set.}. In essence, the unstable manifold consists of the most long-lived trajectories that eventually escape from the cavity (for forward propagation in time), and for this reason is closely related to the high-$Q$ modes. Ray tracing simulations of the unstable manifold have been used successfully to predict for example the far-field intensity distributions of wave-chaotic cavities, and here we extend this method for the first time to the intra-cavity and emission intensity distributions. It should be noted that the most general and frequent case of partially chaotic and regular (i.e., mixed) ray dynamics presents additional complications and is beyond the scope of this article. 

We calculate the unstable manifold of the chaotic saddle using the so-called sprinkler method \cite{Schneider2002}. A large ensemble of trajectories with unit intensity that are distributed uniformly in phase space are launched and their evolution inside the billiard as well as the decay of their intensity due to refractive escape is calculated as a function of time. The algorithm is further explained in \ref{sec:algoRayTrace}. 

The decay of the total intensity $I(t)$ of the trajectories remaining inside the billiard is shown in figures~\ref{fig:spectraDecay}(c) and \ref{fig:spectraDecay}(d) for the D-cavity and the stadium, respectively. After an initial transient period, the total intensity in both cavities decays exponentially with $I(t) \propto \exp(-t / \taucl)$. We call $\taucl$ the classical lifetime\footnote{The inverse, $1 / \taucl$, is called natural decay rate in \cite{Claus2018}.} since it represents the average lifetime of ray trajectories in the cavity in the long-time limit and thus the maximal lifetime of resonances in the semiclassical limit. 

In this regime, the intensity distribution of the rays inside the billiard becomes conditionally invariant, that is, it no longer changes in time except for an overall exponential decay \cite{Claus2018, Lee2004, Ryu2006a}, and the remaining trajectories represent a good approximation of the unstable manifold of the chaotic saddle. A representation of the unstable manifold in a Poincar\'e surface of section (PSOS) of phase space is shown in \ref{app:PSOS+FF}. 

The intra-cavity and emission intensity distributions during the regime of exponential decay are calculated by averaging over all rays during the time interval $29.7$--$44.5~nR/c$ for the D-cavity ($29.7$--$44.5~na/c$ for the stadium). The exact time interval is not important since the ray distributions are conditionally invariant during the exponential decay regime. However, since the evolution of the ray trajectories between reflections instead of just the reflections at the boundaries needs to be tracked to obtain the intra-cavity intensity distributions, the time interval should be longer than several mean scattering times $\left< t_s \right> = \left< l_s \right> n/c$, where $\left< l_s \right>$ is the average distance between two reflections at the boundary. The mean scattering times are $\left< t_s \right> = 1.341~nR/c$ for the D-cavity and $\left< t_s \right> = 1.091~na/c$ for the stadium (see \ref{sec:cavArea}). In addition, the finite $\mathrm{NA}$ of the imaging system is taken into account to properly compare the simulated emission profiles with the measured ones. More details are given in \ref{sec:algoRayTrace}. 

\section{Comparison of experiments, wave and ray simulations} \label{sec:comp}

\subsection{Classical lifetimes and thresholds}
First, we consider the lifetimes of the high-$Q$ modes and the lasing thresholds. The classical lifetime predicted by the ray tracing simulations is $\taucl = 6.3507~\neff R / c$ ($\taucl = 7.1736~\neff a /c$) for the D-cavity (stadium) as shown in \reffig{fig:spectraDecay}. The predicted classical quality factor, $\Qcl = k c \taucl$ where $c$ is the speed of light in vacuum, for a D-cavity with $R = 20~\mu$m (stadium with $a = 23.8~\mu$m) at $\lambda = 800$~nm is $\Qcl^\mathrm{(D)} = 3362$ ($\Qcl^\mathrm{(Stad)} = 4519$). The classical lifetimes are compared to the calculated resonance wavenumbers in figures~\ref{fig:spectraDecay}(a) and \ref{fig:spectraDecay}(b), where the corresponding values of $\Im(k) = -1 / (2 c \taucl)$ are indicated as dashed horizontal lines. 

Since the ray trajectories remaining in the cavity converge to the set of most long-lived trajectories in time, their lifetime $\taucl$ should be an upper limit for the lifetimes of the high-$Q$ modes of the resonator in the semiclassical limit \cite{Harayama2015}. For the same reason, the intensity distribution of this set of ray trajectories should predict those of the high-$Q$ modes. For the D-cavity, $\taucl$ is indeed in good agreement with the lifetimes of the most long-lived modes [see \reffig{fig:spectraDecay}(a)], with only a few modes slightly exceeding $\taucl$. In the case of the stadium shown in \reffig{fig:spectraDecay}(b), however, there are several modes with clearly longer lifetimes than $\taucl$. Such modes are hence called supersharp resonances \cite{Novaes2012}. 

The existence of supersharp resonances is explained by a higher degree of localization compared to other high-$Q$ modes \cite{Casati1999b, Novaes2012}, for example due to scarring (i.e., localization) \cite{Heller1984, Gmachl2002, Harayama2003} on short unstable periodic orbits (UPOs) confined by total internal reflection. However, since the average scarring strength decreases in the semiclassical limit of large resonators \cite{Vergini2012, Vergini2015}, the lifetimes of supersharp resonances converge to $\taucl$ in this limit \cite{Novaes2012}. This has been verified numerically for the stadium billiard \cite{Cerjan2019}, and we do not expect any supersharp resonances for the five to ten times larger cavities investigated experimentally. It is interesting to note that there are no supersharp resonances for the D-cavity in \reffig{fig:spectraDecay}(a), which does not feature modes strongly scarred on long-lived orbits either. 

Experimentally we observe that the stadium cavities have consistently lower lasing thresholds than the D-cavities with the same area. For the cavities presented in \reffig{fig:spectraLI}, the threshold of the stadium is about $1.3$ times lower than that of the D-cavity, and the average over several D-cavities with $R = 100~\mu$m and stadia with $a = 119~\mu$m yields a ratio of $1.25$ \cite{Cerjan2019}. This agrees with the ray tracing simulations that predict a longer classical lifetime for the stadium, but the ratio of $\Qcl^\mathrm{(Stad)} / \Qcl^\mathrm{(D)} \simeq 1.34$ is somewhat higher than the ratio of the measured thresholds. In practice, additional loss mechanisms such as surface roughness play a role. While the quality of the cavity sidewalls is very good (see \cite{Bittner2018a, Cerjan2019}), surface roughness is not negligible. Since surface roughness tends to reduce the difference in the $Q$-factors for different cavity shapes, it can explain the slightly lower ratio of the thresholds of D-cavities and stadia found experimentally. 

\subsection{Interior intensity distributions}

\begin{figure*}[tb]
\begin{center}
\includegraphics[width = 15 cm]{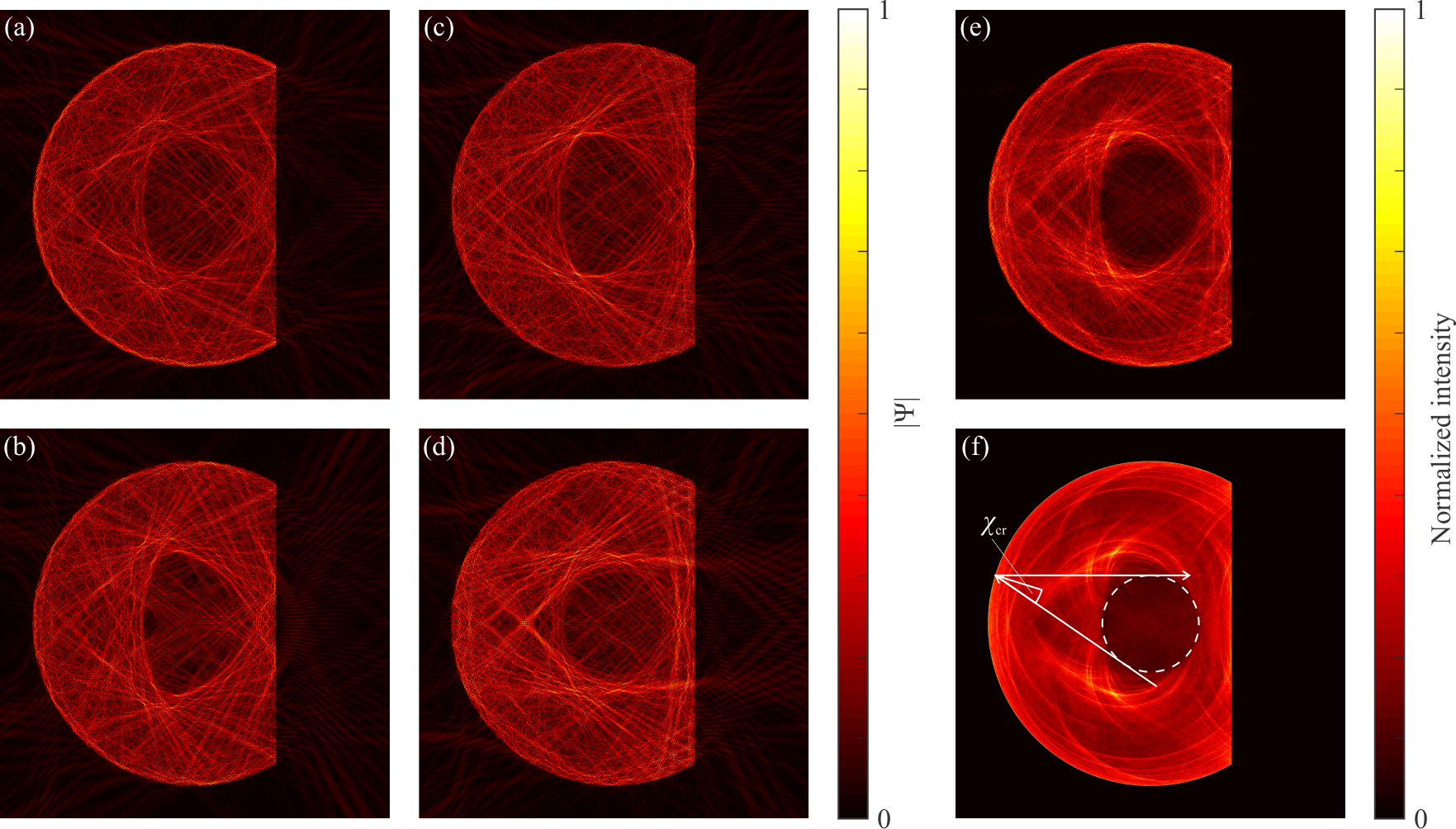}
\end{center}
\caption{Wave simulations of the intensity distributions in a D-cavity with $R = 20~\mu$m and $\neff = 3.37$. (a)~Mode with $\Re(kR) = 157.01$ and $Q = 3531$, (b)~mode with $\Re(kR) = 157.07$ and $Q = 3699$, (c)~mode with $\Re(kR) = 157.15$ and $Q = 3170$, and (d)~mode with $\Re(kR) = 157.22$ and $Q = 3434$. These four modes are indicated by labels a--d in \reffig{fig:spectraDecay}(a). (e)~Sum of the intensity distributions of the $11$ modes with $Q \geq 0.9~\Qcl$. (f)~Ray tracing simulation of the interior intensity distribution. The dashed white circle indicates the caustic with radius $R / \neff$, which results from trajectories reflected at the circular boundary with incident angle equal to the critical angle for total internal reflection, $\chicrit$. The color scale corresponds to $|\Psi|$ in panels (a)--(d) and to the intensity $|\Psi|^2$ in panels (e) and (f).}
\label{fig:IID-Dcav}
\end{figure*}

\begin{figure*}[tb]
\begin{center}
\includegraphics[width = 15 cm]{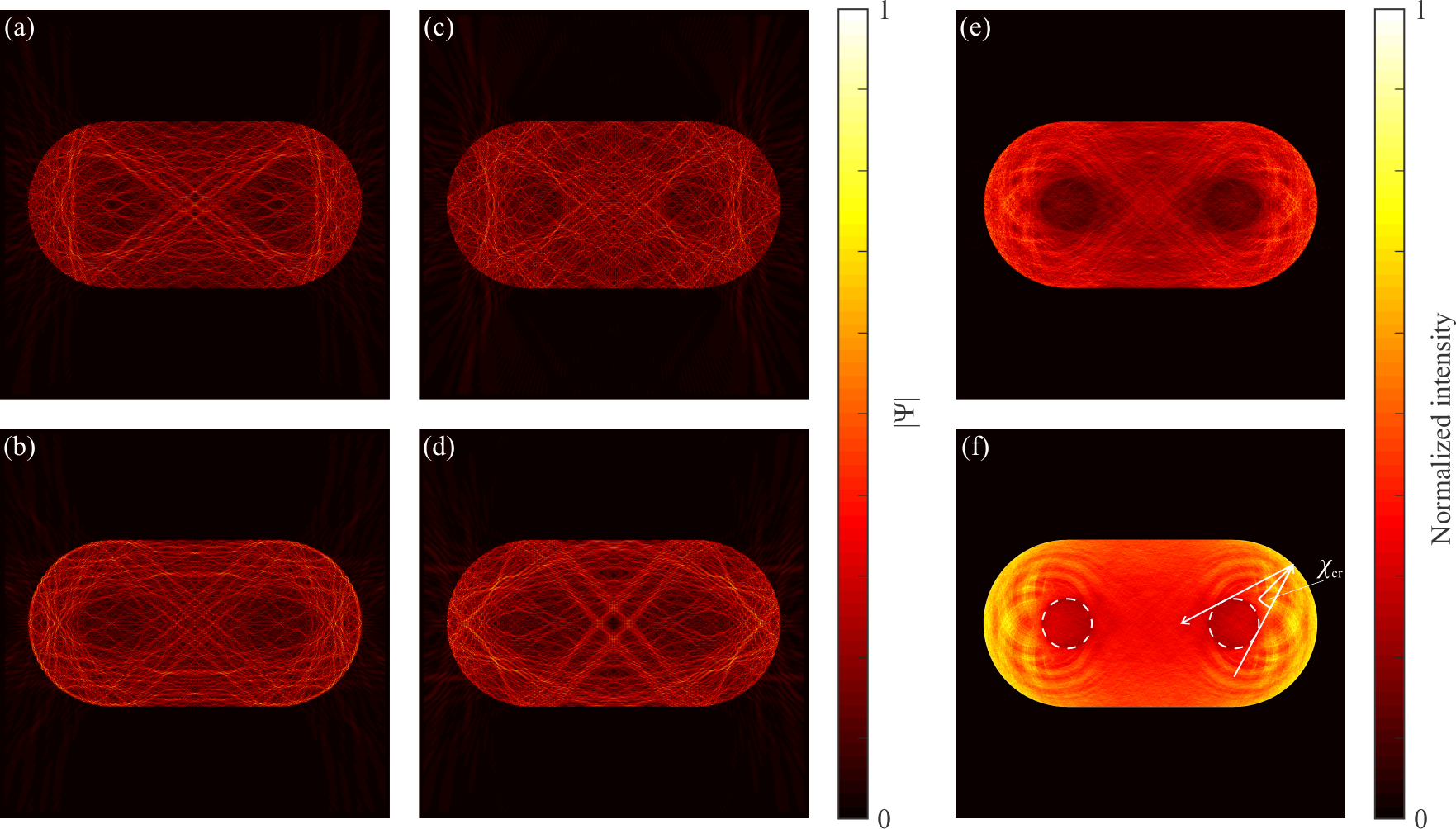}
\end{center}
\caption{Wave simulations of the intensity distributions in a stadium with $a = 23.8~\mu$m and $\neff = 3.37$. (a)~Mode with $\Re(ka) = 186.81$ and $Q = 5855$, (b)~mode with $\Re(ka) = 186.89$ and $Q = 6962$, (c)~mode with $\Re(ka) = 186.99$ and $Q = 4605$, and (d)~mode with $\Re(ka) = 187.02$ and $Q = 5431$. These four modes are indicated by labels a--d in \reffig{fig:spectraDecay}(b). (e)~Sum of the intensity distributions of the $24$ modes with $Q \geq 0.9~\Qcl$. (f)~Ray tracing simulation of the interior intensity distribution. The dashed white circles indicate the caustics with radius $a / (2 \neff)$. The caustics result from trajectories reflected at the circular arcs with the incident angle equal to the critical angle for total internal reflection, $\chicrit$. The color scale corresponds to $|\Psi|$ in panels (a)--(d) and to the intensity $|\Psi|^2$ in panels (e) and (f).}
\label{fig:IID-Stad}
\end{figure*}

Next we consider the interior intensity distributions and compare the ray tracing with the wave simulations. The intra-cavity intensity distributions of several high-$Q$ modes of the D-cavity and the stadium are shown in figures~\ref{fig:IID-Dcav}(a)--(d) and figures~\ref{fig:IID-Stad}(a)--(d), respectively. While their fine structure is clearly different, their coarse structure shows common features. In particular, the modes of the D-cavity in figures~\ref{fig:IID-Dcav}(a)--(d) feature a roughly circular region in the center of the cavity with significantly lower intensity. Similarly, two circular regions of low intensity at the centers of the circular arcs are found for the modes of the stadium in figures~\ref{fig:IID-Stad}(a)--(d). 

We add the intra-cavity intensity distributions of high-$Q$ modes because experimentally we observe the total emission from all lasing modes. The modes are added in intensity since the lasing modes are phase incoherent. The summation highlights the common features of the intensity distributions of individual modes, and better agreement with ray tracing simulations is obtained \cite{Choi2008, Shinohara2008, Shinohara2009, Harayama2015}. We sum the field intensities of all modes with $Q \geq 0.9~\Qcl$, which are $11$ modes for the D-cavity and $24$ modes for the stadium, respectively. These modes are highlighted by the gray areas in figures~\ref{fig:spectraDecay}(a) and \ref{fig:spectraDecay}(b). The threshold of $0.9~\Qcl$ was chosen since $\Qcl$ gives the scale of what can be considered to be a high-$Q$ mode for a given cavity geometry, but the results do not depend sensitively on the chosen threshold. The intensity distributions are normalized such that the integral over the interior of the cavity, $\int_S |\Psi(x, y)|^2 dA$, is equal to $1$ before adding them. 

The total intra-cavity intensity distributions are presented in \reffig{fig:IID-Dcav}(e) and \reffig{fig:IID-Stad}(e) for the D-cavity and the stadium, respectively. The circular regions of low intensity are even more pronounced in the sum of intensity distributions. Even outside these regions, the total intensity distributions are not homogeneous and exhibit points and lines of high intensity. The intra-cavity intensity distributions resulting from the ray-tracing simulations are shown in \reffig{fig:IID-Dcav}(f) and \reffig{fig:IID-Stad}(f), respectively, and show very good agreement with the total intensity distributions of the calculated modes. The circular regions of low intensity as well as many other features are accurately predicted by the ray tracing simulations. 

The most prominent feature of the intra-cavity intensity distributions, the circular regions of lower intensity, are naturally explained by the classical ray dynamics. Trajectories with an angle of incidence larger than the critical angle for total internal reflection, $\chicrit$, are obviously among the most long-lived orbits and contribute significantly to the unstable manifold. The trajectories that hit the circular boundaries exactly at the critical angle form a caustic with radius $R / \neff$ indicated by the dashed white circles in figures~\ref{fig:IID-Dcav}(f) and \ref{fig:IID-Stad}(f). They delimit precisely the regions of low intensity, and it was checked that the radius of the low-intensity regions scales indeed as $1 / \neff$ as a function of the refractive index $\neff$. Even though no experimental data for the intra-cavity intensity distributions are available for comparison, the good agreement of ray and wave simulations of the intensity distribution validates the ray tracing approach. Moreover, the structure of the intra-cavity distributions and in particular the caustics allow us to better understand the emission intensity distributions discussed in the following. 

\subsection{Emission intensity distributions}

\begin{figure*}[tb]
\begin{center}
\includegraphics[width = 12 cm]{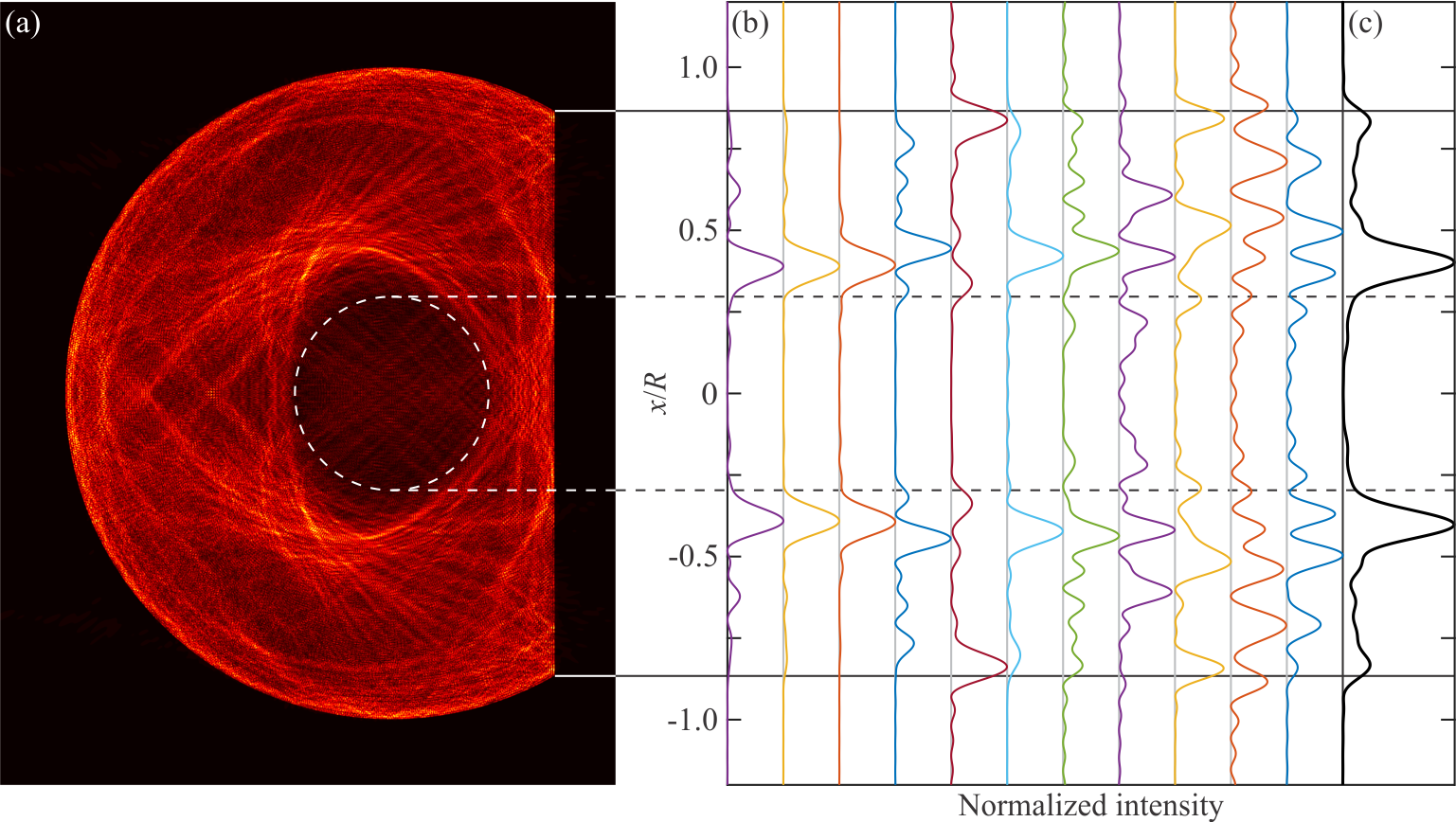}
\end{center}
\caption{Wave simulations of intra-cavity intensity distributions and emission intensity distributions at the straight sidewall for $\mathrm{NA} = 0.4$ of a D-cavity with $R = 20~\mu$m and $\neff = 3.37$. (a)~Sum of the intensity distributions of the $11$ modes with $Q \geq 0.9~\Qcl$ [cf.\ \reffig{fig:IID-Dcav}(e)]. The white dashed circle indicates the caustic with radius $R / \neff$. (b)~Emission intensity distributions of the $11$ modes with $Q \geq 0.9~\Qcl$. The curves are horizontally offset. (c)~Total emission intensity distributions of the $11$ modes. The solid horizontal lines indicate the corners of the cavity and the dashed horizontal lines the projection of the caustic.}
\label{fig:SWEID-Dcav-sim}
\end{figure*}

Wave simulations of the emission intensity distributions at the straight sidewall of a D-cavity are shown in \reffig{fig:SWEID-Dcav-sim} together with the sum of the interior intensity distributions for all modes with $Q \geq 0.9~\Qcl$. The emission intensity distributions are obtained from the calculated wave functions by applying a Fourier filter with width corresponding to $\mathrm{NA} = 0.4$ in order to account for the imaging optics (cf.\ \cite{Bittner2018a}). Figure~\ref{fig:SWEID-Dcav-sim}(b) shows the emission profiles of the $11$ individual modes. While there are clear differences between their intensity distributions, all of them exhibit a region of low intensity in the center, and most have peaks at the border of the center region and near the corners at $x = \pm \sqrt{3} R / 2$. These common features are responsible for the structure of the total emission intensity distribution in \reffig{fig:SWEID-Dcav-sim}(c), which features a region of low intensity surrounded by two peaks and two further peaks near the corners as observed experimentally. It is due to the common features of the modes that the coarse structure of the measured intensity distributions is independent of the pump current and the active lasing modes. 

The region of low intensity in the center results from the caustic with radius $R / \neff$, shown as dashed white circle in \reffig{fig:SWEID-Dcav-sim}(a). The projection of the caustic is indicated by the horizontal dashed lines in \reffig{fig:SWEID-Dcav-sim} and delimits the center region of low intensity quite precisely. The caustic can be seen so clearly in the emission intensity distributions because rays impinging on the straight sidewall with near normal incidence contribute most to the emission distribution due the finite angle of collection of the objective with $\mathrm{NA} = 0.4$, and there are practically no long-lived rays that go through the caustic and hit the straight segment perpendicularly. Figure~\ref{fig:SWEID-Dcav-sim}(a) also shows that the caustic is surrounded by regions with high intensity above and below, which explains the two peaks surrounding the center region in the emission profiles. In the same way, the features with relatively high intensity near the corners in \reffig{fig:SWEID-Dcav-sim}(a) explain the peaks of emission intensity at the two ends of the straight segment. Similar arguments apply to the emission intensity distributions of the stadium which also exhibit a region of low intensity in the middle even though it is not as pronounced as for the D-cavity. 

Next we compare the emission intensity distributions from ray and wave simulations with the experimentally measured ones. Figures~\ref{fig:sweid}(e), \ref{fig:sweid}(f) and \ref{fig:sweid}(g) show the emission profiles measured experimentally well above threshold, calculated by ray tracing and obtained from wave simulations, respectively. The ray and wave simulations show very good agreement with the measured profiles for both cavity geometries. It should be noted that the smallest feature sizes of the emission intensity distributions from wave calculations with a width approximately given by the resolution limit, $\lambda / (2 \, \mathrm{NA})$, appear broader than those of the measured intensity distributions because the distributions are presented as function of the transverse coordinate normalized by the cavity size, and the experimental cavities are five times larger. 

Finally we compare the top view emission intensity distributions just outside of the cavities and at the rings surrounding the cavities with the ray tracing simulations. Figures~\ref{fig:topViewNF}(a) and \ref{fig:topViewNF}(b) show the top view emission images around a D-cavity and a stadium respectively, and figures~\ref{fig:topViewNF}(c) and \ref{fig:topViewNF}(d) the corresponding ray simulations of the intensity just outside the cavities. These images highlight from which parts of the boundary the emission originates and are in good agreement. Figures~\ref{fig:MidFieldDistr}(e) and \ref{fig:MidFieldDistr}(f) show the emission scattered at the rings surrounding the D-cavity and stadium, respectively. Measurements and ray simulations agree well, demonstrating that also the emission directions of the cavities are accurately predicted by our simulations.  

\section{Summary and outlook} \label{sec:conc}
 
The excellent agreement of the ray tracing results with the experimental data and passive cavity wave simulations confirms that the ray simulations can predict precisely the intensity distributions inside and outside the D-cavity and stadium microlasers. Because the ray and wave simulations --- which do not take nonlinear light-matter interaction into account --- agree so well with experiments, we conclude that the nonlinear interaction of modes and active medium does not have a perceivable influence on the structure of the intensity distributions, in contrast to the case of broad-area Fabry-Perot lasers \cite{Mehuys1987, Abraham1990, Lang1991}. Also the influence of surface roughness appears to be negligible, and the inhomogeneous intensity distributions are dictated by the geometry of the cavities alone. 

The ability of ray tracing simulations to comprehensively predict the intra-cavity and exterior intensity distributions as well as classical lifetimes of cavities with fully chaotic ray dynamics is very useful to estimate, e.g., the emission directionality, the lasing thresholds or the strength of modal interaction. They hence provide a computationally efficient tool for the design of asymmetric microcavity lasers, in particular in the semiclassical limit where cavities are often too large for a numerical solution of the Helmholtz equation. However, ray tracing simulations can only predict the intensity distribution of a sum of many lasing modes, not of individual modes. Furthermore, the cavity needs to be much larger than the wavelength so that the high-$Q$ modes are indeed based on the unstable manifold of the chaotic saddle as assumed for the ray tracing simulations. Especially for smaller cavities, significant variations from mode to mode can occur \cite{Choi2008}, and in particular when strong scarring on UPOs is involved \cite{Fang2007b}. 

More work is needed to understand the regime of validity of the ray tracing predictions as far as the classical lifetimes are concerned. In the semiclassical limit, the lifetimes of the fully chaotic stadium and D-cavities are predicted by the time constant $\taucl$ of the exponential decay of the rays remaining in the cavities. Exponential decay in ray tracing simulations was also observed in Refs.~\cite{Lee2004, Ryu2006a, Shinohara2007}, and good agreement of $\taucl$ with the lifetimes of the high-$Q$ modes was demonstrated in \cite{Shinohara2007}. However, chaotic dielectric billiards do not always exhibit exponential decay, and we observe non-exponential decay for D-cavities with a smaller section cut off (i.e., a cut farther away from the center). This is attributed to the existence of families of marginally unstable periodic orbits (MUPOs) \cite{Altmann2013, Dettmann2009}. For smaller cuts than at $R/2$, the D-cavities feature an increasingly large family of equilateral triangle orbits as well as other polygonal MUPO families. Since these orbits are confined by total internal reflection, they and nearby trajectories contribute significantly to the long-term decay of the system. For the D-cavity with cut at $R/2$ considered here, however, the only MUPO family consists of the orbits along the diameter, which have a short lifetime due to their normal incidence at the cavity boundary. Similarly, for the stadium the only MUPO family consists of the so-called bouncing ball orbits between the straight sides of the stadium which are very short-lived as well. In conclusion, for the D-cavity and the stadium geometries considered here, the MUPO families have no contribution to the exponential decay of the rays remaining inside the cavities. In general, the estimation of the classical lifetime for fully chaotic billiards with long-lived MUPO families will not be as straightforward as for the cases considered here. 

Because the ray tracing simulations only take geometric effects and refractive escape into account, they predict that the $Q$-factors scale linearly with the linear cavity size. The very good agreement of the ray simulations with the experimental results shows that refractive escape is indeed the dominant decay mechanism for the optical field in the cavities considered here. In general, however, various wave effects can also contribute significantly to the losses, and their importance depends sensitively on the ratio of cavity size to wavelength. Hence a quantitative understanding of all possible effects involved requires further studies. For example, scarring can result in unusually long-lived resonances \cite{Novaes2012, Casati1999b} and interference between scar contributions from different UPOs may yield significant fluctuations of the $Q$-factors as a function of cavity size and geometry \cite{Fang2005a, Fang2007a}. However, the scarring strength decreases in the semiclassical limit \cite{Vergini2012, Vergini2015}, and also the contributions from other wave effects such as tunneling loss at curved interfaces diminish for increasingly large cavities. 

While we consider only cavities with fully chaotic ray dynamics here, most asymmetric microcavities feature mixed ray dynamics, i.e., they exhibit both chaotic and integrable regions in phase space. Dielectric billiards with mixed dynamics usually feature non-exponential decay of the ray trajectories inside \cite{Ryu2006a}. Moreover, contributions from dynamical tunneling between the integrable and chaotic regions of phase space need to be taken into account to predict the lifetimes of their resonances \cite{Baecker2009, Gehler2015}. Therefore, while ray tracing simulations have proven themselves as a powerful tool to understand and predict the properties of asymmetric microcavities, a complete understanding of the physical mechanisms that determine the intensity distributions and lifetimes for arbitrary cavity geometries and accurate predictions based on semiclassical methods remain an important future challenge. 

\begin{acknowledgments}
The authors thank A.\ D.\ Stone, M.\ Constantin, O.\ Hess and T.\ Harayama for fruitful discussion. The work at Yale is supported partly by the Office of Naval Research (ONR) with MURI grant N00014-13-1-0649, and the Air Force Office of Scientific Research (AFOSR) under grant FA9550-16-1-0416. Y.~Z.\ and Q.~J.~W.\ acknowledge support from the Ministry of Education, Singapore (grants MOE2016-T2-1-128 and MOE2016-T2-2-159) and the National Research Foundation, Competitive Research Program (NRF-CRP18-2017-02).
\end{acknowledgments}

\appendix

\section{Cavity area and mean free path length} \label{sec:cavArea}
The area of the D-cavity shown in \reffig{fig:cavGeom}(a) is $S = R^2 ( 2 \pi / 3 + \sqrt{3} / 4) \simeq 2.527 \, R^2$ and its circumference $\partial S = R (4 \pi / 3 + \sqrt{3}) \simeq 5.921 \, R$. This yields an area of $S = 25,274~\mu\mathrm{m}^2$ for $R = 100~\mu$m. The stadium shown in \reffig{fig:cavGeom}(b) has an area of $S = a^2 (1 + \pi/4)$ and a circumference of $\partial S = a (2 + \pi)$, yielding an area of $S = 25,283~\mu\mathrm{m}^2$ for $a = 119~\mu$m. The mean free path length in a two-dimensional billiard with ergodic dynamics is given by (see \cite{Chernov1997} and references therein)
\begin{equation} \left< l_s \right> = \frac{\pi S}{\partial S} \, . \end{equation}
This yields $\left< l_s \right> \simeq 1.341 \, R$ for the D-cavity and $\left< l_s \right> \simeq 1.091 \, a$ for the stadium. 

\section{Ray-tracing algorithm} \label{sec:algoRayTrace}
For the ray tracing simulations, an ensemble of $10^7$ trajectories are launched with random initial positions uniformly distributed inside the cavity (not just at the boundary) and random initial directions uniformly distributed in the interval $[0, 2 \pi)$. All rays start with unit intensity, and their intensity is reduced according to the Fresnel coefficients for $\neff = 3.37$ and p-polarization at each reflection. The actual time of flight is tracked (instead of just the number of reflections) as recommended in \cite{Altmann2013}. 
	
The intensity profiles inside and outside of the cavities are calculated by sampling the positions and intensities of all rays during the time interval $[29.7$--$44.5]~nR/c$ ($[29.7$--$44.5]~na/c$) for the D-cavity (stadium) with a spatial resolution of $10^{-3}~R$ ($10^{-3}~a$). The far-field intensity distributions and the intensity distributions at the ring surrounding the cavities are sampled analogously as a function of the azimuthal angle $\varphi$. When calculating the emission intensity distributions, only rays with $|\sin(\alpha)| \leq \mathrm{NA}$ are considered, where $\alpha$ is the angle of the outgoing ray with respect to the optical axis of the imaging optics and $\mathrm{NA} = 0.4$ is the numerical aperture of the objective used in the experiments. 

\section{Phase space and far-field distributions} \label{app:PSOS+FF}

\begin{figure}[tb]
\begin{center}
\includegraphics[width = 8.4 cm]{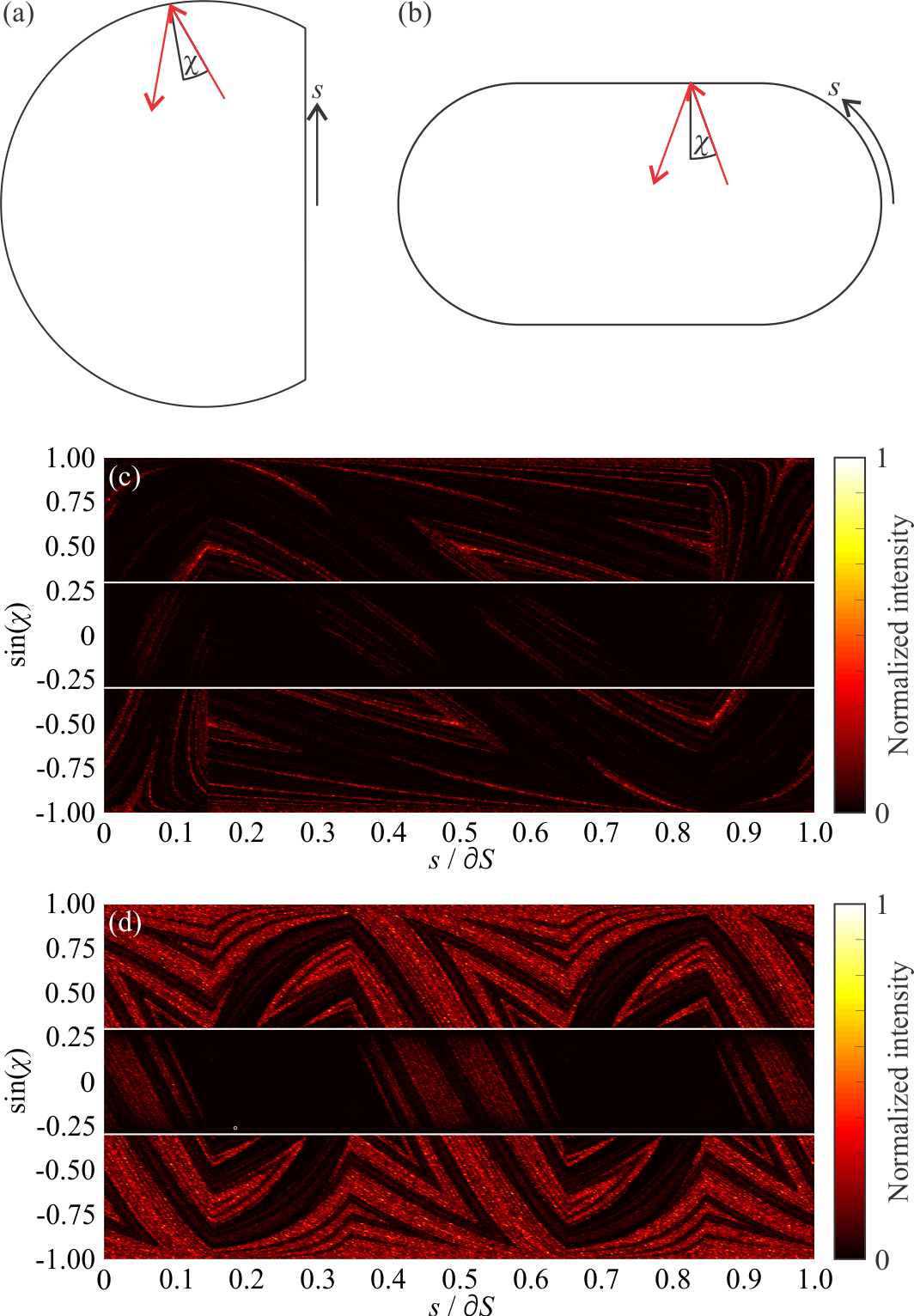}
\end{center}
\caption{Birkhoff coordinates for (a)~D-cavity and (b)~stadium, where $s$ is the position along the cavity boundary and $\chi$ is the angle of incidence on the boundary. The unstable manifold of the chaotic saddle for (c)~the D-cavity and (d)~the stadium for TE polarization and refractive index $\neff = 3.37$. The intensity in the leaky region where rays escape refractively, delimited by the horizontal white lines, is increased by a factor of two for better visibility.}
\label{fig:PSOS}
\end{figure}

The unstable manifold of the chaotic saddle in a PSOS of phase space is shown in \reffig{fig:PSOS}. The PSOS is parametrized in the so-called Birkhoff coordinates \cite{Birkhoff1927} defined in figures~\ref{fig:PSOS}(a) and \ref{fig:PSOS}(b), where $s$ is the position along the boundary of the billiard and $\chi$ the angle of incidence. The boundaries of the lossy region in phase space with $|\sin(\chi)| < 1 / \neff$ are indicated by the horizontal white lines. Only the reflections in the lossy region contribute to the emission intensity distributions, and the intensity in it is significantly lower compared to the regions confined by total internal reflection, $|\sin(\chi)| \geq 1 / n$. 

\begin{figure}[tb]
\vspace{1 cm}
\begin{center}
\includegraphics[width = 8.4 cm]{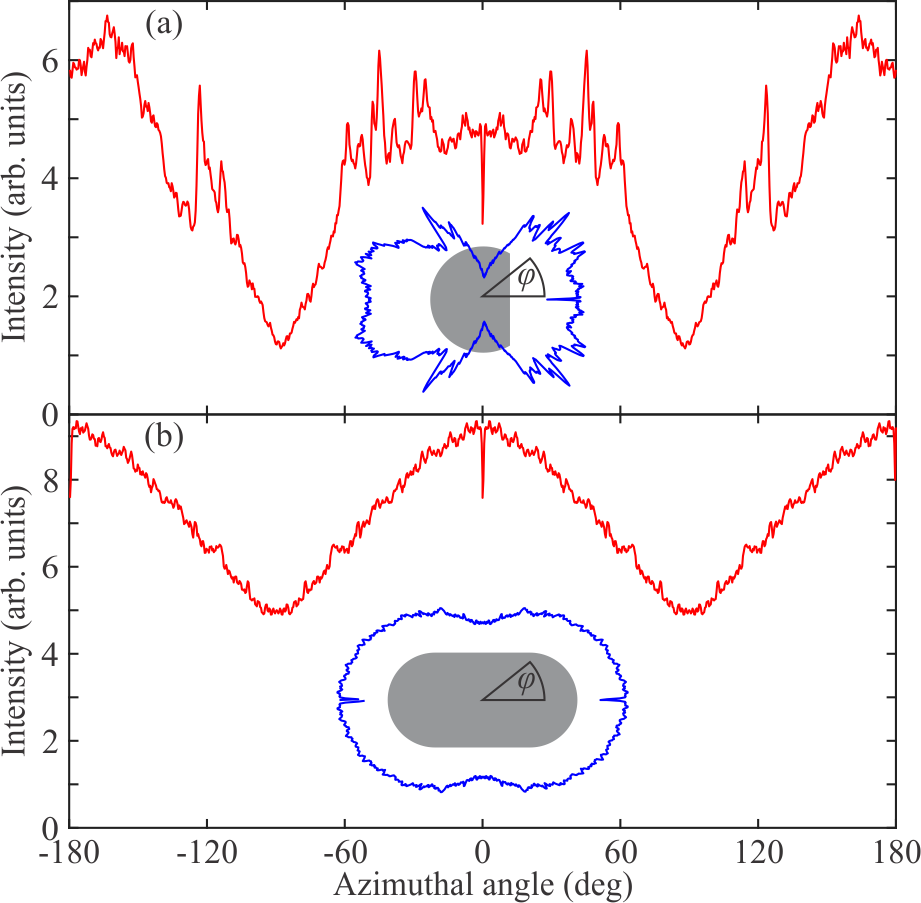}
\end{center}
\caption{Far-field intensity distributions from ray tracing simulations of (a) D-cavity and (b) stadium for TE polarization and refractive index $\neff = 3.37$. The insets (blue lines) show the far-field intensity distributions in polar coordinates and the definition of the azimuthal far-field angle $\varphi$.}
\label{fig:farfield}
\end{figure}

The ray tracing calculations of the far-field intensity distributions are shown in \reffig{fig:farfield}. While the far-field distributions are not uniform since the cavities are asymmetric, their emission is far from being directional, and both D-cavity and stadium exhibit significant emission into almost all directions. This can be partially attributed to the fact that the unstable manifolds shown in \reffig{fig:PSOS} cover a fairly large part of the leaky region and thus contribute to many different emission angles. 

%

\end{document}